\documentclass[apl,twocolumn,preprintnumbers,amsmath,amssymb]{revtex4}

\usepackage{graphicx}% Include figure files
\usepackage{bm}% bold math
\usepackage{color}
\usepackage{amsmath}
\usepackage{amssymb}

\begin{document}

\title{Propagation of Exchange Bias in CoFe/FeMn/CoFe Trilayers}
\author {D. N. H. Nam}
%\thanks{On leave from Institute of Materials Science, VAST, Vietnam}
\email{daonhnam@yahoo.com}
\author {W. Chen}
\author {K. G. West}
\author {D. M. Kirkwood}
\author {J. Lu}
\author {S. A. Wolf}
\affiliation{Department of Materials Science and Engineering, University of Virginia, Charlottesville, VA 22904, USA}
\date{\today}% It is always \today, today,
\begin{abstract}
CoFe/FeMn, FeMn/CoFe bilayers and CoFe/FeMn/CoFe trilayers were grown in magnetic field and at room temperature. The exchange bias field $H_{eb}$ depends strongly on the order of depositions and is much higher at CoFe/FeMn than at FeMn/CoFe interfaces. By combining the two bilayer structures into symmetric CoFe/FeMn($t_\mathrm{FeMn}$)/CoFe trilayers, $H_{eb}^t$ and $H_{eb}^b$ of the top and bottom CoFe layers, respectively, are both enhanced. Reducing $t_\mathrm{FeMn}$ of the trilayers also results in enhancements of both $H_{eb}^b$ and $H_{eb}^t$. These results evidence the propagation of exchange bias between the two CoFe/FeMn and FeMn/CoFe interfaces mediated by the FeMn antiferromagnetic order.
\end{abstract}
\maketitle
%
%\section{Introduction}
%

The magnetization loop, $M(H)$, of the ferromagnetic (FM) layer in a ferromagnetic/antiferromagnetic (FM/AF) thin film structure can be shifted from zero field if it was grown or cooled in the presence of an external magnetic field. This phenomenon, called exchange bias or unidirectional exchange anisotropy, was originally discovered on partially oxidized Co particles by Meiklejohn and Bean more than 50 years ago \cite{Meiklejohn1,Meiklejohn2}. Today, exchange biased ferromagnets are widely used as a key component in spintronic devices \cite{Wolf} such as spin-valves and magnetic tunneling junctions. Despite having been long discovered, widely used in practical applications and under intense investigations, the underlying physics of this intriguing effect still remains open to debate.

The exchange bias phenomenon has been generally considered as an interfacial phenomenon, implying that only interfacial spins are responsible for the unidirectional pinning of the magnetization of the FM layer. However, there have been a number of experimental evidences for an important role the AF bulk effect may play in exchange bias. For instance, the exchange bias field $H_{eb}$ was found to be strongly dependent on the thickness of the AF layer even for large thicknesses \cite{Ambrose,Sang,Li}, or exchange bias can persist even when a layer of nonmagnetic spacer such as Cu was inserted in between the AF and FM layers \cite{Gokemeijer,Gruyters}. In FM/AF/FM trilayer structures, a bulk characteristic of exchange bias would imply that exchange bias should propagate from one AF/FM (or FM/AF) interface to the other. However, recent studies on trilayers gave different results depending on material specifics and magnetic heat treatments. Using a field cooling procedure at the plateau field that separates the minor switching loops of the two FM layers in Py/FeMn/Co trilayers, Yang and Chien \cite{Yang} observed that the top and bottom exchange bias systems are coupled via a spiraling spin structure across the intervening FeMn layer. Leung \textit{et al.} \cite{Leung1} later suggested that such a macroscopic AF spin spiral was due to the specific field cooling treatment employed in the work, but not a universal feature of exchange bias trilayers. While no sign of bias propagation was observed in NiFe/FeMn/Co trilayers grown in a low field ($\sim$5 Oe) \cite{Leung2}, it was seen in those that were field cooled in 1 kOe from above the blocking temperature \cite{Leung1}. Blamire \textit{et al.} \cite{Blamire} reported that no propagation of spin order from the bottom biased Co/FeMn to the top FeMn/CuNi interfaces was observed in Co/FeMn/CuNi structures grown in a magnetic field $H$=200 Oe.

Our work on bilayers shows that there is a large difference in exchange bias between FeMn/CoFe and CoFe/FeMn bilayer structures due to the influence of the magnetized CoFe on the establishment of the AF order of FeMn. Remarkably, we have also observed that adding a CoFe seed layer to the FeMn/CoFe bilayer significantly improves the bias of the top CoFe. On the other hand, deposition of a top CoFe on the CoFe/FeMn bilayer enhances the bias field of the bottom CoFe layer. The top and bottom CoFe layers in trilayers both show an enhanced bias with decreasing the sharing FeMn layer thickness. These results support the presence of significant propagation of exchange bias in the FeMn layer in CoFe/FeMn/CoFe trilayers.
%
%\section{Experiment}
%

The CoFe/FeMn, FeMn/CoFe bilayers and symmetric CoFe/FeMn/CoFe trilayers were grown at room temperature (in an \textit{in-situ} magnetic field of 50 Oe applied along the substrate surface) using a Biased-Target Ion Beam Deposition (BTIBD) technique \cite{Zhurin,Hylton}. The base pressure of the BTIBD system was $\sim$2$\times$10$^{-7}$ Torr, the Ar processing pressure was $\sim$7$\times$10$^{-4}$ Torr, the target bias voltage was kept at 600 V for all the depositions, and the CoFe and FeMn target compositions are Co$_{95}$Fe$_5$ and Fe$_{50}$Mn$_{50}$, respectively. All the samples were grown on Si wafers, whose surface was covered by a seed layer of Ta(5) (for CoFe/FeMn bilayers and trilayers) or Cu(5) (for FeMn/CoFe bilayers), and capped by a Ta(5) layer except for CoFe/FeMn bilayers that use a cap of 5 nm of Cu (all the thickness units are in nm). The use of Cu layers was aimed at promoting the $\gamma$-fcc FeMn phase required for exchange bias. The samples subject to field cooling were heated (from room temperature) to 250 $^\mathrm{o}$C in 4.5 minutes and then furnace cooled to room temperature in a field of 3 kOe; the whole heating and cooling process was performed in a flowing mixture of N$_2$+5\% H$_2$. The exchange bias fields, $H_{eb}$=$|$$H_{c1}$+$H_{c2}$$|$/2, were measured at 305 K by a Quantum Design PPMS-6000 for all the samples in both as-deposited and field-cooled states. Here, $H_{c1}$ and $H_{c2}$ are the coercivities determined on the opposite field sweeping directions that are measured at $M/M_s$=0 for the bilayers and at $M/M_s$=$\pm$0.5 separately for the two CoFe layers in the trilayers.
%
%\section{Results and Discussion}
%
\begin{figure}[t!]
%   Requires \usepackage{graphicx}
%  \vspace{0.1in}
  \includegraphics[width=7.5cm]{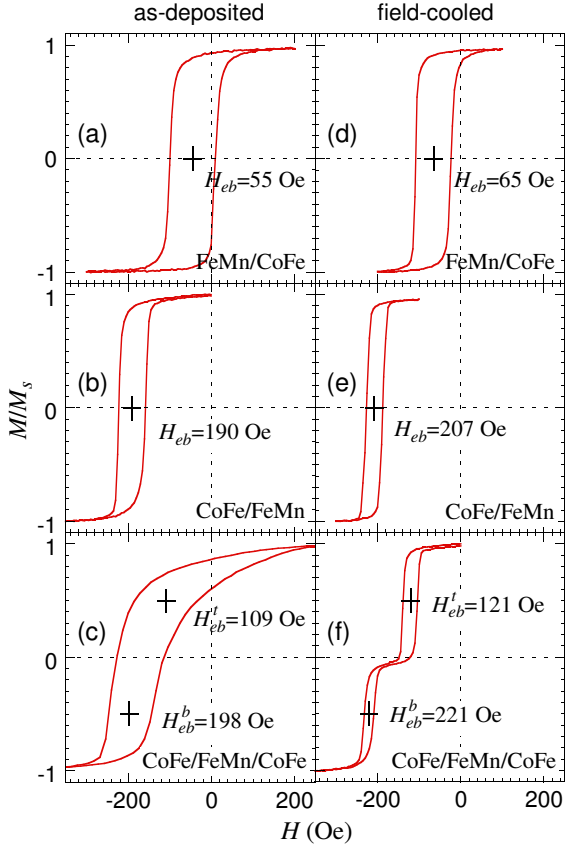}
  \caption{(Color online) $M(H)$ loops of the [(a), (d)] Si/Cu(5)/FeMn(10)/CoFe(4)/Ta(5), [(b), (e)] Si/Ta(5)/CoFe(4)/FeMn(10)/Cu(5) bilayers, and [(c), (f)] Si/Ta(5)/CoFe(4)/FeMn(10)/CoFe(4)/Ta(5) trilayers. The samples were grown in $H$=50 Oe. (a), (b), and (c) are for as-deposited samples; (d), (e), and (f) are for field-cooled samples. The crosses mark the center of the main or minor $M$($H$) loops where $H_{eb}$ is determined.}\label{Fig1}
\end{figure}
\begin{figure}[t!]
%   Requires \usepackage{graphicx}
%  \vspace{0.1in}
  \includegraphics[width=7.5cm]{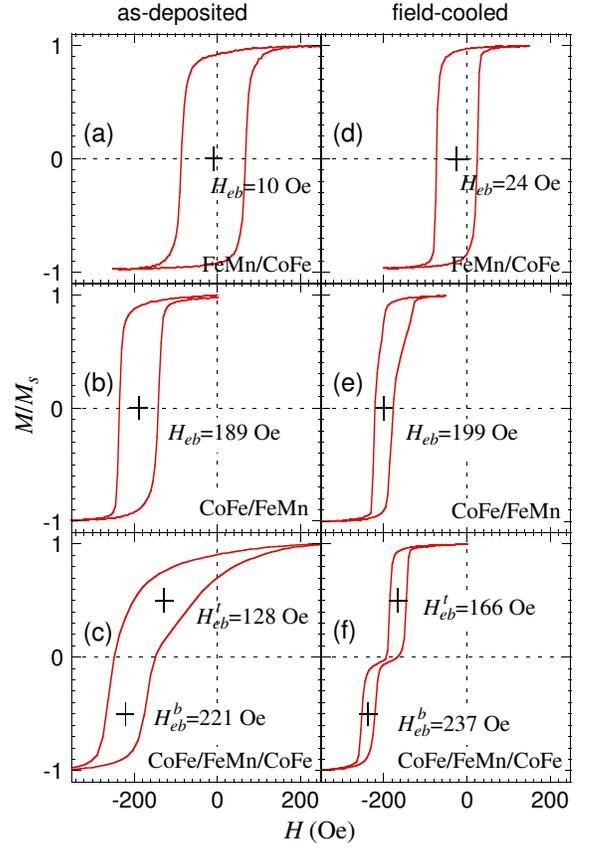}
  \caption{(Color online) $M(H)$ loops of the [(a), (d)] Si/Cu(5)/FeMn(6)/CoFe(4)/Ta(5), [(b), (e)] Si/Ta(5)/CoFe(4)/FeMn(6)/Cu(5) bilayers, and [(c), (f)] Si/Ta(5)/CoFe(4)/FeMn(6)/CoFe(4)/Ta(5) trilayers. The samples were grown in $H$=50 Oe. (a), (b), and (c) are for as-deposited samples; (d), (e), and (f) are for field-cooled samples. The crosses mark the center of the main or minor $M$($H$) loops where $H_{eb}$ is determined.}\label{Fig2}
\end{figure}

Typical $M(H)$ data of our bilayers and trilayers are presented in Fig. \ref{Fig1} for FeMn(10)/CoFe(4), CoFe(4)/FeMn(10), and CoFe(4)/FeMn(10)/CoFe(4) structures grown in $H$=50 Oe. As for as-deposited bilayers, we have commonly observed that exchange bias is much stronger in CoFe/FeMn than in FeMn/CoFe structures. That trend is clearly demonstrated in Figs. \ref{Fig1}(a) and 1(b) where a large difference in $H_{eb}$ between FeMn(10)/CoFe(4) ($H_{eb}$=55 Oe) and CoFe(4)/FeMn(10) ($H_{eb}$=190 Oe) bilayers is observed. This effect seems to be caused by changes in micromagnetic structure due to magnetic interactions rather than in the crystalline structure and texture of the FeMn layer associated with the underlayer effect, since a seed Cu layer was used for the FeMn/CoFe structure. In the case of CoFe/FeMn, because of a strong polarizing field produced by the CoFe surface magnetization that competes against the formation of the AF order, uncompensated spins are easily created when FeMn is grown on the saturated CoFe in magnetic field. On the other hand, depositing CoFe on a stable FeMn layer may have a much smaller affect on its well established AF order, resulting in a significantly weaker exchange bias. A much higher \textit{in-situ} field may be required to create as strong an exchange bias in FeMn/CoFe as that in CoFe/FeMn bilayers. Within this context, the concentration of uncompensated spins may be one factor that determines the exchange coupling between AF and FM layers across their interface.

For the trilayers, by varying the thickness of the CoFe layers, we can easily identify the minor loops for the top and bottom CoFe layers. The hysteresis loop of the as-deposited CoFe(4)/FeMn(10)/CoFe(4) trilayer in Fig. \ref{Fig1}(c) shows only a slight increase in the exchange bias field $H^b_{eb}$ of the bottom CoFe(4) layer (lower minor loop; $H^b_{eb}$=198 Oe) in comparison to that of the corresponding CoFe/FeMn bilayer [$H_{eb}$=190 Oe, Fig. \ref{Fig1}(b)]. In contrast, a huge change is induced in the top CoFe layer. The broad magnetization reversals of the top layer (upper minor loop)  indicates that its exchange coupling with the FM layer is not uniform; a major part of the layer seems to switch its magnetic moment synchronically with the bottom one. The nominal exchange bias field $H^t_{eb}$ measured at $M/M_s$=0.5 is 109 Oe, which is much higher than the value of 55 Oe obtained for the corresponding FeMn/CoFe bilayer [Fig. \ref{Fig1}(a)]. These results unambiguously indicate that exchange bias is strongly improved in the top CoFe layer in the presence of the bottom CoFe layer. It is unlikely that the bottom CoFe layer would cause a larger improvement of the $\gamma$-fcc AF phase in the FeMn layer than by a Cu underlayer. On the other hand, the presence of the top CoFe is not expected to make any change in the crystalline structure of the FeMn layer underneath. Therefore, the improvements of exchange bias of both CoFe layers must be indicative of a magnetic coupling between the two CoFe/FeMn and FeMn/CoFe systems sharing the same intervening FeMn layer.

Nevertheless, the increase of 8 Oe (or 4\%) in exchange bias field of the bottom CoFe layer from bilayer [Fig. \ref{Fig1}(b)] to trilayer [Fig. \ref{Fig1}(c)] is somewhat too small to be conclusive. If the top and bottom exchange bias systems are magnetically coupled, $H^t_{eb}$ and $H^b_{eb}$ would increase with decreasing the FM layer thickness, $t_\mathrm{FeMn}$. Fig. \ref{Fig2} plots the hysteresis loops of similar bilayer and trilayer structures but with a thinner (6 nm) FeMn layer. As expected, with decreasing $t_\mathrm{FeMn}$ from 10 to 6 nm, $H^t_{eb}$ increases from 109 to 128 Oe and $H^b_{eb}$ from 198 to 221 Oe [compare Figs. \ref{Fig1}(c) and \ref{Fig2}(c)]. Moreover, $H^b_{eb}$ of the $t_\mathrm{FeMn}$=6 nm trilayer is now 32 Oe (or 17\%) higher than that of the corresponding CoFe layer in bilayer. It is worth noting here that, while dipolar (or "orange-peel"-type) coupling between the two CoFe layers may not be avoidable in our trilayers, that cannot be responsible for the increase of $H^b_{eb}$ with decreasing $t_\mathrm{FeMn}$. Our data (not shown here) indicate that, due to the dipolar coupling, $H^b_{eb}$ starts to decrease when $t_\mathrm{FeMn}$ is decreased to below 6 nm (e.g., $H^b_{eb}$=204 and 173 Oe for $t_\mathrm{FeMn}$=5 and 4 nm, respectively). A huge increase in exchange bias is also obtained for the top CoFe layer in trilayer [Fig. \ref{Fig2}(c)] with reference to the bilayer [Fig. \ref{Fig2}(a)]. It is very interesting that as $t_\mathrm{FeMn}$ decreases, while exchange bias fields are reduced in both top and bottom CoFe bilayers, they are strongly increased in the corresponding trilayers. All of these facts convincingly suggest that there exists a mutual propagation of exchange bias between the top and bottom interfaces through the intermediate FeMn. Even for $t_\mathrm{FeMn}$ of up to 25 nm, the top CoFe layer exchange bias is still induced by the bottom one. Although the shape of the $M(H)$ loops of the trilayers in Figs. \ref{Fig1}(c) and \ref{Fig2}(c) look rather similar to those observed by Yang and Chien \cite{Yang}, our angular measurements indicate that both the CoFe layers have the same easy axis as that initially created by the deposition field, thus avoiding any possibility of a spiraling magnetic structure in these samples.

As shown in Figs. \ref{Fig1}(d)-(f) and \ref{Fig2}(d)-(f), it is surprising that field cooling the samples from 250 $^\mathrm{o}$C and in $H$=3 kOe only slightly improves their exchange bias fields, indicating that the exchange bias states established in the as-deposited samples were already close to equilibrium. The biggest change is observed for the top CoFe layer in both the FeMn(10) [Fig. \ref{Fig1}(f)] and FeMn(6) [Fig. \ref{Fig2}(f)] trilayers, where exchange bias becomes uniform and also improved, indicating some sort of redistribution of uncompensated spins caused by the field cooling process. Qualitatively, the behaviors of the field-cooled samples are in general the same as that observed for the as-deposited ones. We varied the field-cooling temperature from 180 to 300 $^\mathrm{o}$C and observed no significant change in our results. The large difference in exchange bias between field-cooled CoFe/FeMn and FeMn/CoFe bilayers (and between the top and bottom interfaces in trilayers as well) is a striking feature that would imply that the field cooling may have reset the coupling of the CoFe layers and uncompensated spins without significantly increasing their concentration. It is therefore possible that a 3 kOe cooling field is still far from enough to bring about equal exchange bias fields for the two exchange bias systems whether in bilayers or trilayers.
%
%\section{Conclusion}
%

In summary, our results have shown that uncompensated spins are created favorably when an antiferromagnet is deposited on a magnetized FM layer. Although the uncompensated spins are created near the FM/AF interface, they also spread over the AF layer to the top AF/FM interface, where creation of uncompensated spins is less favored, leading to a strong improvement of exchange bias of the top FM layer. On the other hand, the top FM layer may also contribute a certain amount of uncompensated spins, resulting in an increase of the bottom FM layer exchange bias. Our results here underline the important of the concentration, as well as the distribution, of uncompensated spins in exchange bias systems and demonstrate that there exists a propagation of exchange bias within the AF layer in FM/AF/FM trilayer structures.

%\begin{acknowledgments}

This work is supported by DMEA under Contract No. H94003-08-2-0803
and ONR under Contract No. N00014-06-1-0428.

%\end{acknowledgments}

%
\end{document}